\documentclass[12pt]{article}

\usepackage{graphicx}
\usepackage{hyperref} 
\usepackage{verbatim} 
\usepackage{url}
\usepackage{epstopdf}
\usepackage[normalem]{ulem}
\usepackage{color}
\usepackage{ulem}
\usepackage{dsfont}
\usepackage[utf8]{inputenc}
\DeclareUnicodeCharacter{2010}{-}
\usepackage{float}
\usepackage{amsmath}
\usepackage{comment}
\usepackage{breqn}
\usepackage{amssymb}
\usepackage{bbold}
\usepackage{times}
\renewcommand\vec[1]{\ensuremath\boldsymbol{#1}}
\newcommand{\upa}{\uparrow}
\newcommand{\dwa}{\downarrow}
\newcommand{\mda}{\rightarrow}
\newcommand{\nea}{\nearrow}
\newcommand{\sea}{\searrow}

\newenvironment{sciabstract}{%
\begin{quote} \bf}
{\end{quote}}

\newcounter{lastnote}

\title{Dynamics of spin and density fluctuations in strongly interacting few-body systems}

\author
{Rafael Emilio Barfknecht,$^{1,2\ast}$ Angela Foerster,$^{1}$ Nikolaj Thomas Zinner$^{2,3}$\\
\\
\normalsize{$^{1}$Instituto de F{\'i}sica da UFRGS, Av. Bento Gon{\c c}alves 9500, Porto Alegre, RS, Brazil,}\\
\normalsize{$^{2}$Department of Physics and Astronomy, Aarhus University, Ny Munkegade 120, Denmark}\\
\normalsize{$^{3}$Aarhus Institute of Advanced Studies, Aarhus University, DK-8000 Aarhus C, Denmark}\\
\\
\normalsize{$^\ast$To whom correspondence should be addressed; E-mail:  rafael.barfknecht@ufrgs.br}
}

\date{}

\begin{document}
\baselineskip24pt

\maketitle


\begin{sciabstract}

The decoupling of spin and density dynamics is a remarkable feature of quantum one-dimensional many-body systems. In a few-body regime, however, little is known about this phenomenon. To address this problem, we study the time evolution of a small system of strongly interacting fermions after a sudden change in the trapping geometry. We show that, even at the few-body level, the excitation spectrum of this system presents separate signatures of spin and density dynamics. Moreover, we describe the effect of considering additional internal states with SU($N$) symmetry, which ultimately leads to the vanishing of spin excitations in a completely balanced system.
\end{sciabstract}

Simplified quantum models of interacting particles in one dimension have been, for decades, a favorite starting point for theoretical physicists of different fields, such as particle physics and condensed matter \cite{giamarchi_book}. This is mainly due to the possibility of finding exact mathematical solutions for these models, which can provide insight into more complex problems. This is the case, for instance, of paradigmatic one-dimensional models with contact interactions, for which the solutions can be found through the method known as the Bethe ansatz \cite{gaudin_book}. As examples, we have the Lieb-Liniger model \cite{lieb1,lieb2} for bosons and the Gaudin-Yang model for two-component fermions \cite{gaudin,yang_fermions}.

Additional interest has been drawn recently to this area due to the arrival of experimental techniques that allow for the construction of effectively one-dimensional systems of cold atoms in optical lattices \cite{bloch1}. These experiments feature a remarkable degree of control over different parameters, such as the confinement frequency, number of atoms and interactions between them \cite{Bloch2012,Gross995}. While theoretical models with exact solutions usually assume periodic boundary conditions, different approaches have made it possible to study one-dimensional systems of atoms in confining geometries that range from the harmonic trap \cite{brouzos,miguel_bosons,pecak1} to double-wells \cite{miguel_TG,sowinski1,sowinski2,simos}, boxes \cite{pecak2,barfk4} and one-dimensional lattices \cite{cartarius,mikkelsen}. 

An effect of particular interest in this context is the phenomenon known as spin-charge separation, where the degrees of freedom related to the density (``charge", in analogy to electronic systems) and spin components can be described independently. In spinless systems, the quantization of phonon modes in many-body bosonic \cite{stringari,PhysRevA.66.043610} and fermionic \cite{PhysRevA.64.033605} ensembles has been well-established through the hydrodynamic approach. In one-dimensional two-component systems, the separation of density and spin dynamics is captured by the celebrated Tomonaga-Luttinger Liquid (TLL) model \cite{haldane}. Theoretical proposals for the observation of spin-charge separation usually employ the TLL description \cite{recati} or assume weakly interacting many-body Fermi \cite{kollath} or Bose \cite{kleine} mixtures. Recently, the hydrodynamic approach has been also used to study these effects in the presence of temperature \cite{mestyan}.
The strongly interacting regime, however, is where this effect is expected to manifest itself more dramatically, with the freezing of the spin degrees of freedom as compared to the density dynamics \cite{fuchs}. Additionally, many works have explored the possibility of describing a strongly interacting mixed system as an effective spin chain \cite{guan2,guan,deuretz2,artem2,pu1,xiaoling,laird}, where the separation of the spin and density sectors arises naturally in a dynamical context \cite{artem3,pu2}. Such a framework is therefore appropriate for studying this particular phenomenon. 


Experimentally, ultracold atomic setups are good candidates for measuring spin-charge separation in detail. Unlike experiments in condensed matter, ultracold atomic ensembles usually allow for a fine tuning of the interactions and the trapping geometries, as well as a precise control over the number of atoms \cite{jochim2}. While spin-charge separation has been observed in quantum wires \cite{Auslaender88}, measurements with cold atoms are still in an early stage\cite{yang_sf,vijayan}. Therefore, new proposals for the observation of this effect, which take into account the possibilities and limitations imposed by current experiments with cold atoms, are welcome. In this work, we provide such a proposal by presenting an analysis of the dynamics of a strongly interacting few-body system of fermions with SU($N$) symmetry after a sudden change - a {\it quench} - in the trapping potential.  Recently, systems of cold atoms in optical lattices with SU(3) and SU(4) symmetries have been explored theoretically \cite{hofstetter1,hofstetter2}. The correlations of SU($N$) impenetrable systems have also been calculated for an increasing number of internal components and display interesting properties \cite{rigol}.

\begin{figure}
\centering
\includegraphics[width=0.8\textwidth]{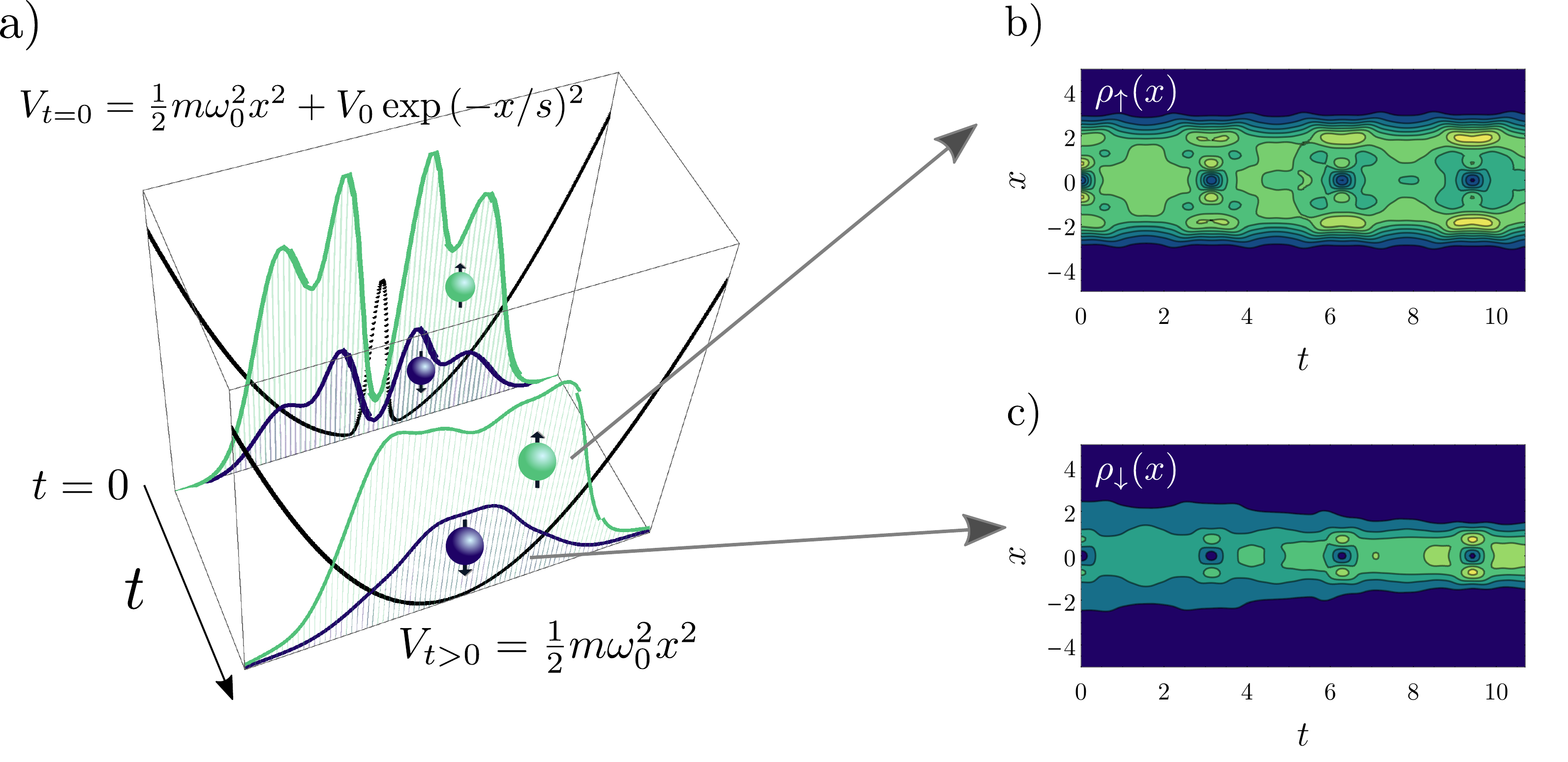}
\caption{a) Sketch of the quench protocol adopted in this work. An imbalanced ($N_\uparrow=3$, $N_\downarrow=1$) strongly interacting two-component system of atoms is initialized in the ground state of a harmonic potential modified by a Gaussian bump centered at $x=0$. For $t>0$ this perturbation is suddenly switched off and the time evolution of the density for each component is registered. The majority (minority) density $\rho_{\uparrow(\downarrow)}(x)$ is shown in green (blue). The black curves, which are rescaled for clarity, show the shape of the trapping potential at $t=0$ and $t>0$. b)-c) Time evolution of the spin densities, where length and time are given in units of the harmonic potential (see text for details). Yellow and blue colors indicate higher and lower densities, respectively. The slices shown in a) correspond to $t/T_0=0$ and $t/T_0=5$ in the figures on the right. This figure has been published, with modifications, in Ref.~\cite{reb_thesis}.}
\label{fig1}
\end{figure}

We begin by presenting the formalism used to describe a system of strongly interacting atoms in a trap, where the Hamiltonian can be mapped into a spin chain with position-dependent exchange coefficients. We then describe the quench protocol, which essentially consists of changing the trap from a split well (where we assume a Gaussian barrier in the center of the system) to a simple harmonic well (see Fig.~\ref{fig1} for a sketch of this protocol). The ground-state configurations for these two systems are considerably different, and by changing the potential we can expect a non-trivial time evolution in the spin and density sectors. Initially, we describe the effect of the quench in the density sector and its consequences on the spin chain dynamics. By combining the dynamics of both sectors, we can extract the signatures of the separation between the density and spin oscillations in the system, showing how this effect can be observed in few-body ensembles, even as the number of internal components is increased. Moreover, we show that for a completely balanced system (where each atom is in a different internal state) the spin signature vanishes, and the excitation spectrum is analogous to that of a gas of impenetrable bosons. Multi-component cold atomic ensembles with strong interactions are currently within experimental reach \cite{fallani} and often exhibit exotic dynamical effects, such as edge states\cite{Stuhl1514,fallani_hall}. In these systems, the internal states of the atoms can be manipulated with laser pulses, and the behavior of each component can be measured with precision. Studying cold atoms with different internal symmetries in a highly controllable environment can lead to insight on particle physics models and even shed light on the validity of unification schemes \cite{pati_salam}. Parts of the work presented here have been published, with modifications, in Ref.~\cite{reb_thesis}.

\section{Results}\label{results}
\paragraph*{System description.}\label{description}

Our goal is to describe the dynamics of a strongly interacting few-body system with internal (``pseudospin") degrees of freedom. We focus our description on a fermionic system, but the formalism is equally valid for bosons with the correct adaptations to the Hamiltonian. We consider initially an SU(2) system, where the internal degrees of freedom are described by $\lvert \upa \rangle$ and $\lvert \dwa \rangle$. Later we will generalize our approach to systems with higher symmetries. We denote the particle numbers in each species by $N_{\upa}$ and $N_{\dwa}$, the total number of particles thus being given by $N=N_{\upa}+N_{\dwa}$. For simplicity, we adopt the notation $N_{\upa}+N_{\dwa}$ (e.g. 3+1 for a system with three particles of species $\lvert \upa \rangle$ and one of species $\lvert \dwa \rangle$). Experimentally, two-component fermionic systems can be realized by preparing trapped ensembles of ${}^{6}$Li atoms in the two lowest hyperfine states \cite{murmann,jochim3}.

The Hamiltonian for the system under consideration is given by 
\begin{equation}\label{hm2}
H=\sum_i^{N} H_0(x_i)+g\sum_{\uparrow,\downarrow}\delta(x_i-x_j)
\end{equation}
where $H_0(x)=-\frac{\hbar^2}{2m}\frac{\partial}{\partial x}+V(x)$ denotes the single particle Hamiltonian, where $V(x)$ is a trapping potential. The remaining term accounts for the contact interactions, where atoms in different internal states interact with strength $g$ (interactions between atoms in the same internal state are forbidden due to the Pauli principle). Since we are dealing with atoms of the same element in different internal states, we consider all masses equal. In our calculations, we use the length, energy and time units of the harmonic trap, that is $l_0=\sqrt{\hbar/m\omega_0}$, $\epsilon_0=\hbar \omega_0$ and $T_0=1/\omega_0$, respectively. The interaction strength $g$ is given in units of $\hbar^2/ml$. We also assume, for simplicity, $\hbar=m=1$.

A case of particular interest in one-dimensional systems is the limit of strong interactions ($g\gg 1$), where the Hamiltonian can be mapped to that of a spin chain \cite{deuretz1,deuretz2,artem2,massignan,pu1,xiaoling}. In the case of fermions, Eq.~\eqref{hm2} can be written, up to linear order in $1/g$, as \cite{laird}
\begin{equation}\label{hamper}
H=E_0-\sum_{i=1}^{N-1}\frac{\alpha_i}{g}(1-P_{i,i+1}),
\end{equation}
where $P_{i,i+1}$ is the permutation operator, which exchanges two neighboring atoms of different components. When all interactions between atoms in different internal states are the same, the system exhibits an SU($N$) symmetry \cite{PhysRevB.45.5293}. In the following sections, we focus only on strong repulsive interactions, and consider a fixed value of $g$ in our calculations. For the formalism considered here, it has been shown that the static properties and spatial distributions are well-described already for $g\sim 10$ \cite{deuretz2}. In the strongly attractive limit, the model described here corresponds to the fermionic analog \cite{fermi_ST} of the Super-Tonks-Girardeau gas \cite{haller1}.

The exchange coefficients $\alpha_i$ are solely determined by the geometry of the trapping potential through the wave function for a system of spinless fermions, which we label $\Phi(x_1,...,x_N)$. Obtaining the eigenstates of Eq.~\eqref{hamper}, we can calculate the spatial distributions of each atomic component in the trap as
\begin{equation}\label{spin densities}
\rho_{\upa,\dwa}(x)=\sum_{i=1}^{N}\rho^{i}_{\upa,\dwa}(x),
\end{equation}
where $\rho^i_{\uparrow,\downarrow}(x)=m^{i}_{\uparrow,\downarrow}\rho^{i}(x)$, $m^{i}_{\uparrow,\downarrow}$ is the probability of finding $(\uparrow,\downarrow)$ spins at site $i$ and $\rho^{i}(x)$ represents the spatial distribution of each individual particle in the trap. In Section~\ref{methods} we provide details on how to calculate the exchange coefficients and the spatial densities.

\begin{figure}
\centering
\includegraphics[width=0.5\textwidth]{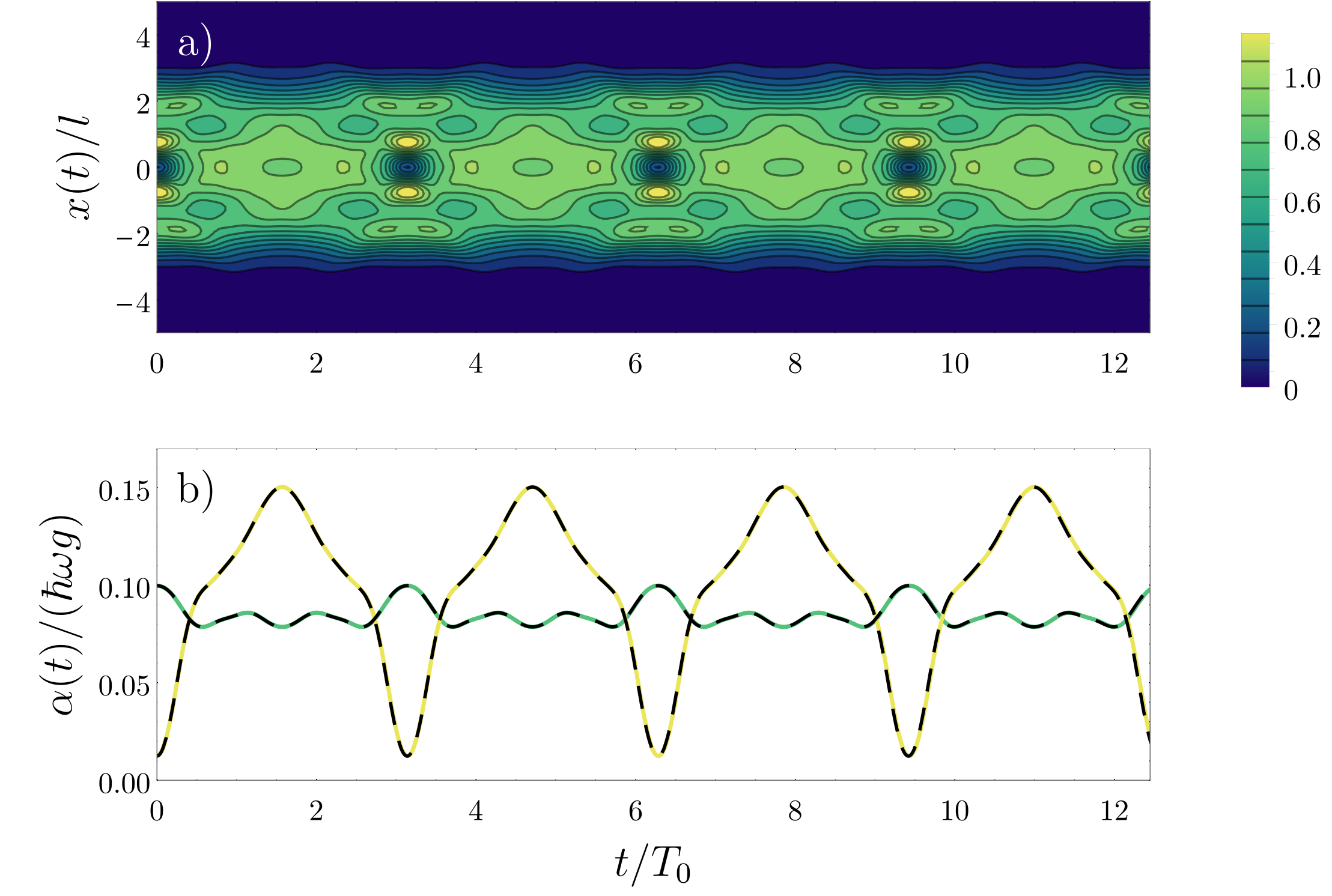}
\caption{Time evolution of a) the total density $\rho_{\text{c}}(x,t)$ and b) the exchange coefficients $\alpha_1(t)$ (green) and $\alpha_2(t)$ (yellow), for the case of $N=4$. Since the trapping potential is spatially symmetric at all times, we have $\alpha_3(t)=\alpha_1(t)$. In b), the black dashed curves show the results obtained with an analytical fit of the data with a Fourier expansion. In b), we assume $g=20$. This figure has been published, with modifications, in Ref.~\cite{reb_thesis}.}
\label{fig2}
\end{figure}

\paragraph*{Dynamics.}
We now describe the procedure that induces the dynamical evolution of the system, which consists of a sudden change of the trapping potential. Our initial choice of $V(x)$ is given by a harmonic trap with an additional Gaussian bump in the center, as shown in Fig.~\ref{fig1} a). For $t>0$, the bump is suddenly turned off. The time evolution of the spinless fermion wave function can then be described in terms of the evolution of the single particle orbitals under the same quench protocol \cite{minguzzi1,cartarius}. We are thus able to construct $\Phi(x_1,...,x_N,t)$ for all times $t$, which in turn determines the time-dependence of the exchange coefficients in Eq.~\eqref{hamper}. As a general rule, we can assume that the exchange coefficients are proportional to the overlap between the single particle distributions; additionally, since the trapping potentials are symmetric at all times, we have $\alpha_i(t)=\alpha_{N-i}(t)$. In Fig.~\ref{fig2} a) we show the time evolution of the spatial densities obtained from $\Phi(x_1,...,x_N,t)$. At this point, we are not considering the spin sector, so the densities shown correspond to the total (``charge") density
\begin{equation}\label{charge}
\rho_{\text{c}}(x,t)=\sum_{i=1}^{N}\rho^i(x,t),
\end{equation}
which is normalized to the total number of particles. Its dynamical behavior is what should be expected for the coherent density oscillations of a Tonks-Girardeau gas \cite{minguzzi1}. In panel b), we see the behavior of $\alpha_1(t)/g$ and $\alpha_2(t)/g$ after the sudden change in the potential. We observe that the periodicity of the dynamics in the motion of the infinitely repulsive gas is reflected in the dynamics of the exchange coefficients. Particularly, since the system is approximately split in two parts at $t=0$, we have that $\alpha(t=0)\sim 0$ (there is nearly no overlap between the initial densities at the center of the system). When the densities in the central region become larger, we see that the numerical value of $\alpha_2(t)$ surpasses that of $\alpha_1(t)$. This indicates that the spin correlations of the system should change between these two points, as we will see next.

Given the established time-periodicity of Hamiltonian.~\eqref{hamper}, we can analyze the dynamics of the spin sector using Floquet theory (see Section \ref{methods} for details). While most studies performed in this context must deal with the issue of thermalization due to the external driving \cite{eckardt1}, in our case the time-dependence of the spin chain originates directly from the dynamics of the density sector generated by the sudden change of the trapping potential.
To describe the time evolution of the system in these terms, we first find an analytical fit of the exchange coefficients in terms of Fourier modes. In Fig.~\ref{fig2} b), this approximation is shown as the black dashed curves. The full time evolution of the spin sector is performed through the numerical integration of the Schr{\"o}dinger equation with the Crank-Nicolson method \cite{crank}. Our quantities of interest are the dynamical densities $\rho_{\text{c}}(x,t)$, $\rho_{\upa}(x,t)$ and $\rho_{\dwa}(x,t)$, as well as the squared width of the distribution for each spin density, defined as $\langle x^{2}_{\upa,\dwa}(t)\rangle=\int dx\,\rho_{\upa,\dwa}(x,t)\,x^2$. While we choose to focus on the time evolution of parity-symmetric operators, other quantities (like, for instance, $\langle x_{\upa,\dwa}(t)\rangle$) could also be explored. Such cases, however, would require a different choice of quench protocol or initial spin state to break the symmetry across the system.

\paragraph*{SU(2).}
We choose initially a fermionic system with SU(2) symmetry. In this case, the permutation operator is given by $P_{i,i+1}=\frac{1}{2}(1+\vec{\sigma_i}\cdot\vec{\sigma_{i+1}})$, which allows us to write the Hamiltonian as
\begin{equation}\label{fermisc}
H=E_0-\frac{1}{2}\sum_{i=1}^{N-1}\frac{\alpha_i}{g}(1-\vec{\sigma_i}\cdot\vec{\sigma_{i+1}}),
\end{equation}
where $E_0$ has the same meaning as in Eq.~\eqref{hamper}. We notice that this Hamiltonian reproduces the fermionic cases presented, for instance, in Refs.~\cite{artem2,artem3}. We focus mostly on the 3+1 case, which can be interpreted as a few-body Fermi background in the presence of an impurity \cite{tylutki}. As we will show next, in this simple setup it is already possible to describe the decoupling of spin and density dynamics. At the end of this section, we include some results for a 4+2 combination, which points to the possibility of realizing the protocol described here in larger systems. 

\begin{figure}[h]
\centering
\includegraphics[width=0.40\textwidth]{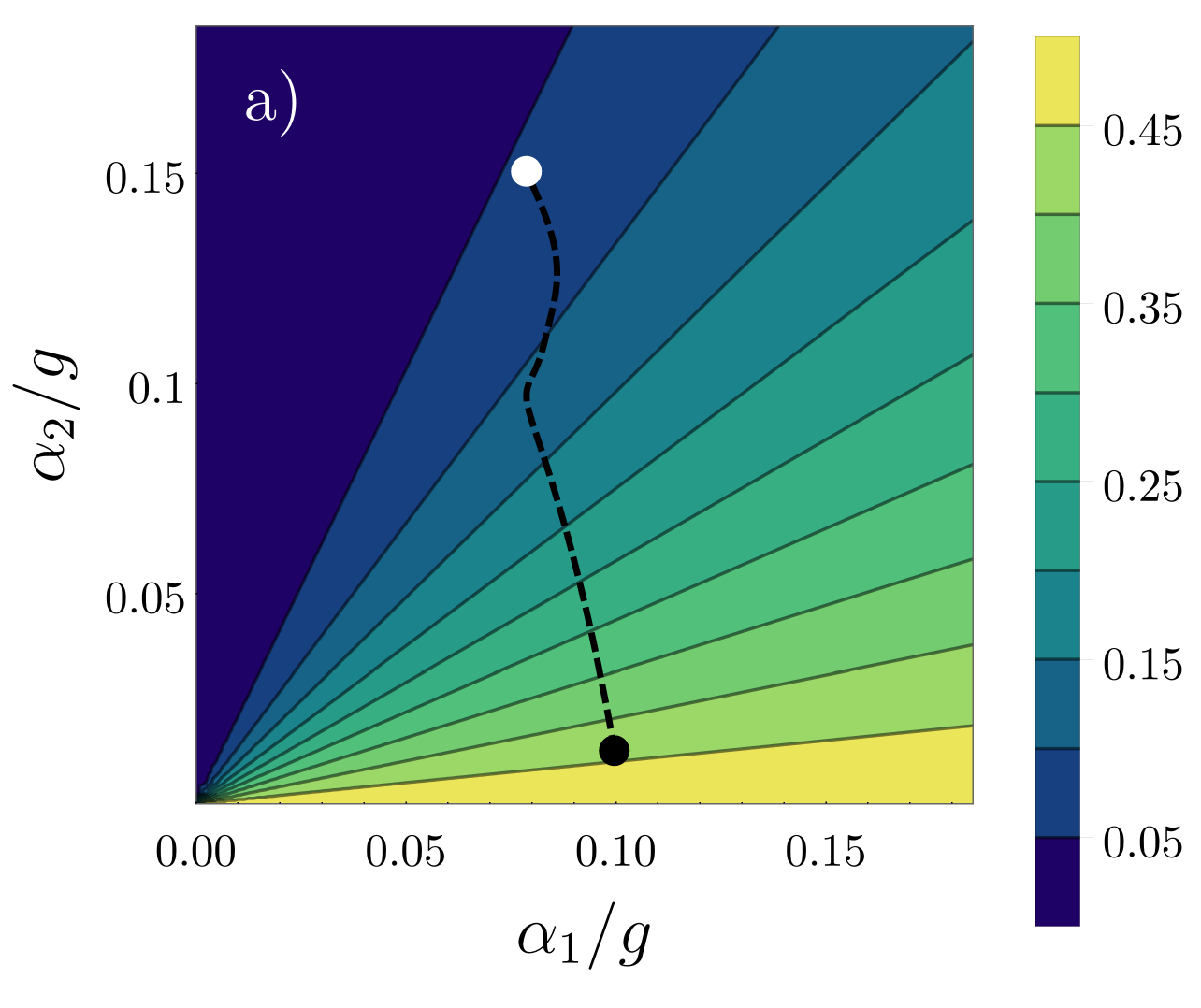}
\caption{Numerical values of $P_{\text{edge}}$ for different combinations of $\alpha_1$ and $\alpha_2$. The black dashed curve corresponds to the trajectory of the exchange coefficients in time after the trap quench. The initial point is marked by the black dot, while the white dot marks the values of the exchange coefficients at $t=T_0/2$. This figure has been published, with modifications, in Ref.~\cite{reb_thesis}.}
\label{fig3}
\end{figure}

The ground state of Hamiltonian \eqref{fermisc} with repulsive interactions ($g>0$) has antiferromagnetic correlations and can be described (in the 3+1 case) by 
\begin{equation}
\lvert \text{gs}\rangle=\lvert \dwa\upa\upa\upa\rangle-\lvert \upa\upa\upa\dwa\rangle+\left(1+\sqrt{2}\right)\left(\lvert \upa\upa\dwa\upa\rangle-\lvert \upa\dwa\upa\upa\rangle\right),
\end{equation}
aside from a normalization factor. Here, we have assumed a homogeneous potential, such that the exchange coefficients are identical and equal to 1. However, it can be shown that similar results hold for a harmonic trap. To observe how the ground state correlations change with the choice of $\alpha_i$, we define the operator $P_{\text{edge}}=\lvert\langle \dwa\upa\upa\upa \rvert \text{gs}\rangle\rvert^2+\lvert\langle \upa\upa\upa\dwa \rvert \text{gs}\rangle\rvert^2$, which provides information regarding the position of the impurity (the $\dwa$ atom) in the system. In Fig.~\ref{fig3} we show the values of $P_{\text{edge}}$ for different choices of $\alpha_1$ and $\alpha_2$, where again we assume a parity-symmetric potential.

We can readily see that, for $\alpha_1=\alpha_2$, we have a constant result of $P_{\text{edge}}=\frac{1}{4}(2-\sqrt{2})$. Above the diagonal ($\alpha_2>\alpha_1$) we have the region that includes, for instance, the coefficients obtained from a harmonic trap. In this case, the antiferromagnetic correlations are even more prevalent. If, however, $\alpha_2\ll 1$, we obtain a larger probability of finding the impurity at the edges. These ground state correlations are obtained when considering a potential such as the double-well assumed for $t=0$. In fact, we can plot the trajectory of the exchange coefficients after the quench in the trapping potential described above. This is shown by the dashed curve in Fig.~\ref{fig3}. The white dot denotes the values at $t=0$; the curve is then traveled back and forth periodically as $t$ increases. The fact that this trajectory crosses over the diagonal indicates that the sudden quench in the potential should induce major changes in the spin correlations of the system.

To quantify this effect we now calculate the time evolution of the spin density and of the squared width for each component. It is important to notice that, for $t>0$, the external potential is a simple harmonic trap. In Fig.~\ref{fig1} b) and c), we show the time evolution of $\rho_{\upa}(x,t)$ and $\rho_{\dwa}(x,t)$, respectively. We notice that, while the underlying dynamics seen in Fig.~\ref{fig2} is still present, we now have an additional oscillation mode. Specifically, after the sudden change in the potential, we observe a tendency of the majority atoms to spread to the edges, while the impurity localizes towards the center.

\begin{figure}
\centering
\includegraphics[width=1.0\textwidth]{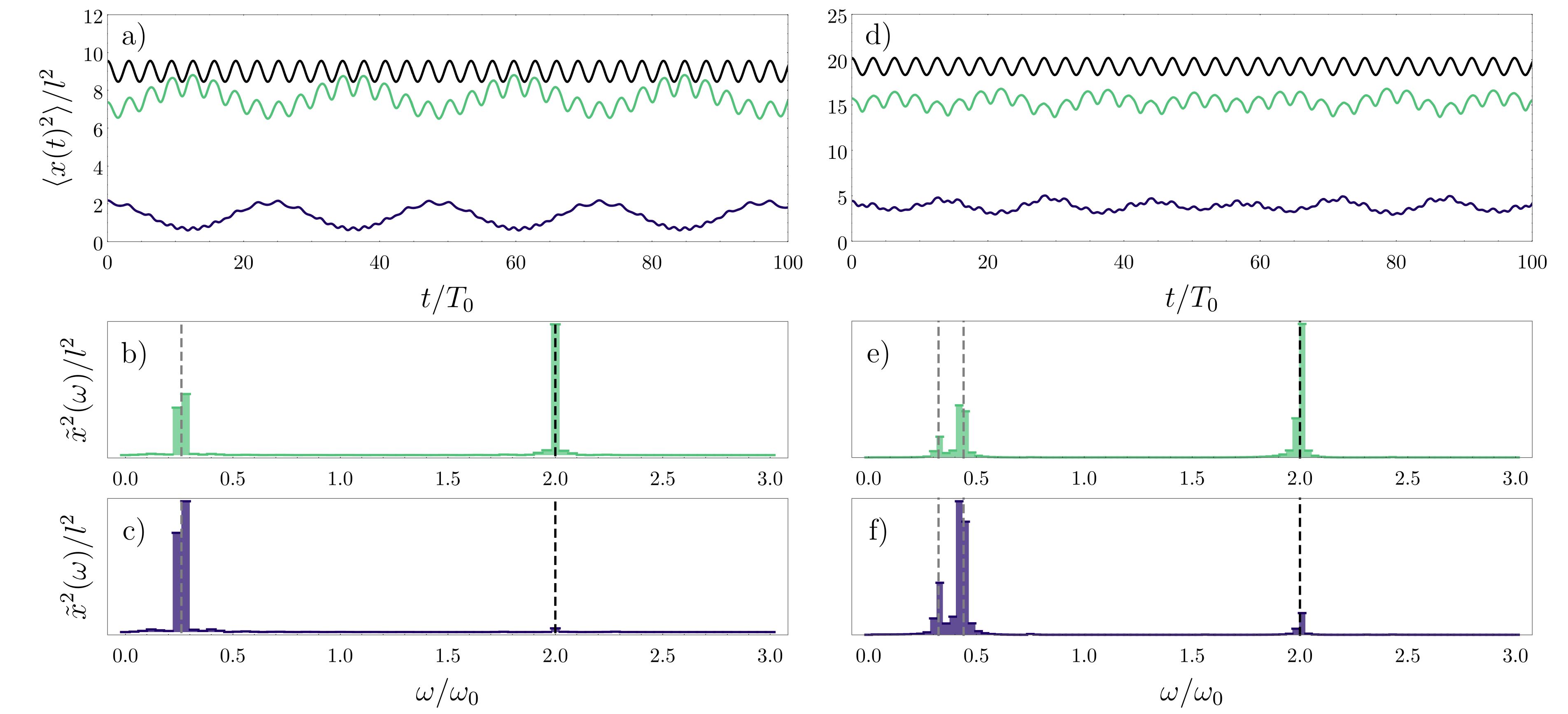}
\caption{Dynamics in the SU(2) fermionic system for the a)-c) $N=4$ (3+1) and d)-f) $N=6$ (4+2) cases. In a) and d) we present the time evolution of the squared widths $\langle x^2_{\uparrow}(t)\rangle$ (green) and $\langle x^2_{\downarrow}(t)\rangle$ (blue) for each spin component, and $\langle x^2_{\text{c}}(t)\rangle$ (black) for the total density. Panels b), c), e) and f) show the excitation peaks in the frequency domain, where the lower frequencies correspond to the spin dynamics. The black dashed lines mark the frequency of the total density oscillations ($\omega=2\omega_0$). The gray dashed lines correspond to gaps in the Floquet quasienergy spectrum of the time-dependent spin chain. In all cases, we assume $g=20$. This figure has been published, with modifications, in Ref.~\cite{reb_thesis}.}
\label{fig4}
\end{figure}

In Fig.~\ref{fig4} a) we show the time evolution of the squared width for the density of each component, over a larger time interval. This can be interpreted as induced breathing  modes for the background and the impurity. Additionally, we show the dynamical behavior of the total density (Eq.~\eqref{charge}). Besides corroborating the results found in Fig.\ref{fig1}b) and c), these curves show how the fluctuations in the density and spin sectors are captured as two oscillations modes in the dynamics of the spin densities for each individual component. In Fig.~\ref{fig4} b) and c) we show the Fourier transform of the width oscillations, defined as $\tilde{x}^2(\omega)=\int dte^{-i \omega t}\langle x^2(t)\rangle$, where the contributions of the spin and density excitations appear as two separate peaks, the lower frequency corresponding to the spin dynamics.

Here, we can see that the dynamics of the minority component is strongly dominated by the spin excitations. On the other hand, the majority component has a more balanced distribution of oscillations in the density and spin sectors. In Fig.~\ref{fig4} b) and c), we additionally include the theoretical predictions for the density and spin oscillations (as black and gray dashed lines, respectively). The first is obtained by calculating the oscillation frequency of the single particle spatial orbitals and corresponds to the expected result for the Tonks-Girardeau or the spinless fermion gas, namely $\omega=2\omega_0$. The second is extracted by calculating the gaps in the Floquet quasienergy spectrum of the time-periodic spin chain. We point out that the coherence in the oscillations of each component over long times is due to the assumption that the spatial sector is described by a spinless fermion wave function, which in a real system is only approximately true for very strong interactions. Depending on the real value of the parameter $g$, we could expect to find excitation peaks which are slightly different than the ones predicted here. In our formalism, modifying $g$ means changing the energy gaps in Eq.~\eqref{fermisc}, which translates into a frequency shift in the spin excitation peaks (with the frequency decreasing as the interaction strength is increased).

In Fig.~\ref{fig4} d)-f), we extend our results to the 4+2 case. The time-dependence of the exchange coefficients for $N=6$ is obtained by calculating these coefficients for one or two periods and then fitting a function by expanding the result in Fourier modes, as done in Fig.~\ref{fig2} for the $N=4$ case. While the general behavior (relative magnitude and position of the excitation peaks) are maintained with respect to the $N=4$ system, we find additional peaks in the low-frequency side of the spectrum, which originates from having a larger number of particles in each component. Nevertheless, these excitations can also be captured by analyzing the Floquet quasienergy gaps of the time-dependent spin chain. These results indicate that the protocol we employ is also suited for larger systems, provided that the imbalance in spin the populations is respected. In the next sections, we show how increasing the number of internal components will affect the behavior of these quantities.

\paragraph*{SU(3).}

We now consider the case of a three-component strongly interacting fermionic gas with SU(3) symmetry. These systems are particularly interesting due to their connections to the quark model in the framework of quantum chromodynamics. We label the three internal states as $\lvert \upa \rangle$, $\lvert \mda \rangle$  and $\lvert \dwa \rangle$. While the Hamiltonian can still be described by Eq.~\eqref{hamper}, the permutation operator is now given by

\begin{equation}\label{permusu3}
P_{i,i+1}=\frac{1}{3}+\frac{1}{2}\vec{\lambda}_i\cdot\vec{\lambda}_{i+1},
\end{equation}
where $\vec{\lambda}$ is the vector composed by the eight generators of the SU(3) group, namely the Gell-Mann matrices. A system described by Eq.~\eqref{permusu3} can be mapped into the Lai-Sutherland model \cite{lai,sutherland} through $P_{i,i+1}=\vec{S}_i\cdot\vec{S}_{i+1}+(\vec{S}_i\cdot\vec{S}_{i+1})^2-1$, which is a particular case of the spin-1 bilinear biquadratic model \cite{schmitt,aguado}.
We keep the number of particles fixed as $N=4$, with $N_\upa=2$, $N_\mda=1$ and $N_\dwa=1$. The quench protocol and the quantities considered are the same as in the previous section.

\begin{figure}
\centering
\includegraphics[width=1.0\textwidth]{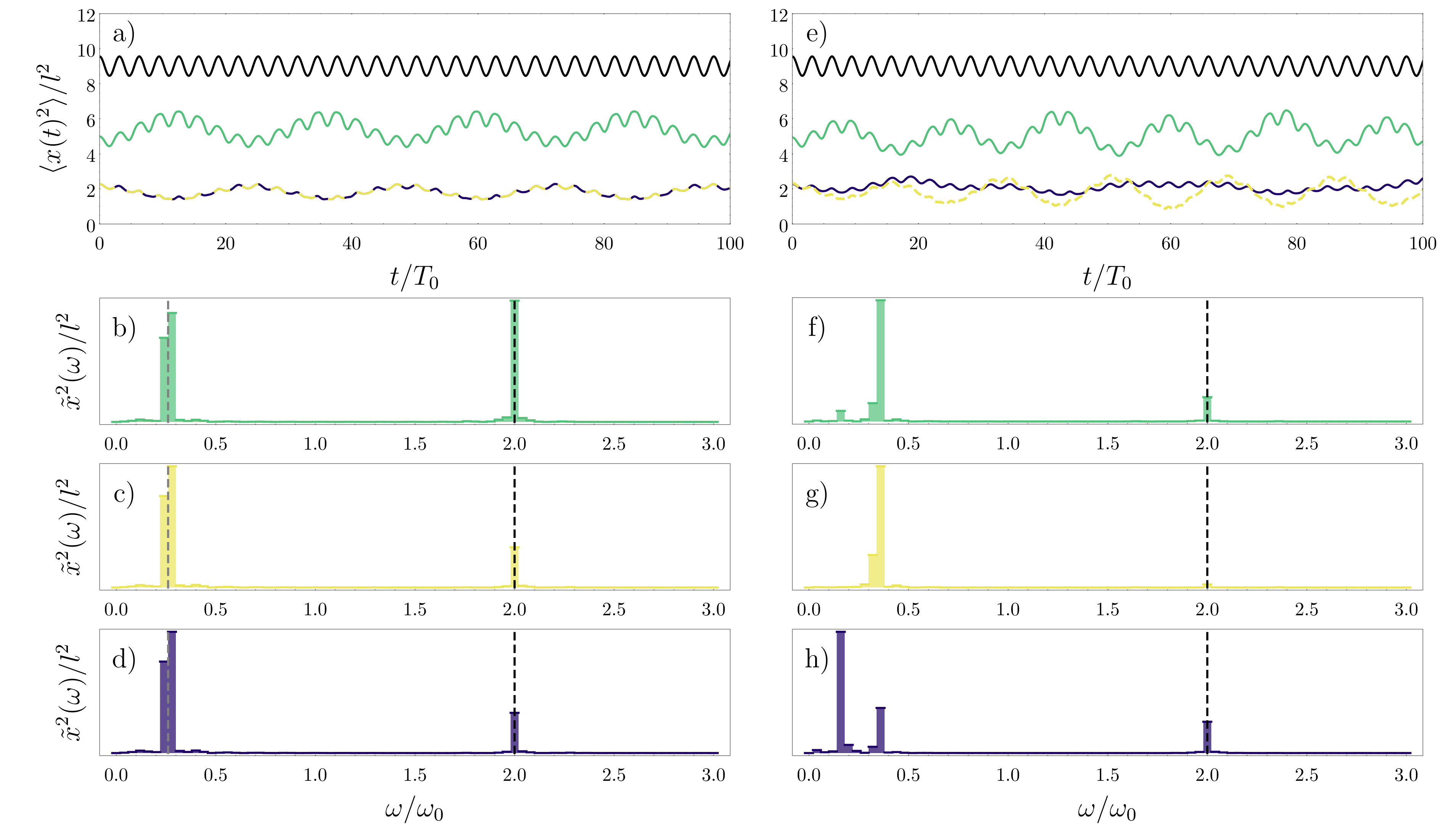}
\caption{Dynamics in the three-component fermionic system. In a)-d) we show the results for the SU(3)-symmetric system, while in e)-h) we present the results when SU(3) symmetry is broken by an anisotropy parameter. In a) and e) we show the time evolution of the density widths $\langle x^2_{\upa}(t)\rangle$ (green), $\langle x^2_{\mda}(t)\rangle$ (dashed yellow), $\langle x^2_{\downarrow}(t)\rangle$ (purple) and $\langle x^2_{\text{c}}(t)\rangle$ (black). Panels b), c) and d) show the excitation peaks in the frequency domain for $\langle x^2_{\upa}(t)\rangle$ and $\langle x^2_{\mda}(t)\rangle$ and $\langle x^2_{\dwa}(t)\rangle$, respectively, while f)-h) show the corresponding results for broken SU(3) symmetry. The black dashed lines mark the frequency of the total density oscillations. The gray dashed lines show the value of the gap in the Floquet quasienergy spectrum of the time-dependent spin chain. In all cases, we assume $g=20$. In e)-h) we have additionally $\eta=1/2$. This figure has been published, with modifications, in Ref.~\cite{reb_thesis}.}
\label{fig5}
\end{figure}

In Fig.~\ref{fig5} a), we show $\langle x(t)^2\rangle$ for each component and for the total density. Since we have two minority particles, each interacting with the remaining atoms with interaction strength $g$, the results for each of these components are identical. The excitation peaks seen in Figs.~\ref{fig5} b), c) and d) reveal the contributions of the density and spin oscillations to the dynamics of each component (naturally, since we have $N_{\mda}=N_{\dwa}=1$ plots c) and d) show identical results). Still, we can see that the majority component has a larger contribution to the total density excitations. The minority cases, however, show a slight increase in these frequencies as compared to the two-component case, with the spin oscillations remaining dominant.

An interesting perspective when dealing with multicomponent strongly interacting gases is a case where interactions are slightly imbalanced and a particular symmetry is broken. Here, we analyze the three-component case with broken SU(3) symmetry. It is useful, in this context, to rewrite the SU(3) permutation operator in terms of raising and lowering operators. These are defined as $T^{\pm}=(\lambda^1\pm i\lambda^2)/2$, $V^{\pm}=(\lambda^4\pm i \lambda^5)/2$ and $U^{\pm}=(\lambda^6\pm i \lambda^7)/2$, where once again $\lambda^i$ are the Gell-Mann matrices. We still focus the particular case of $N=4$, with $N_\upa=2$, $N_\mda=1$ and $N_\dwa=1$.

Below we rewrite the permutation operator with these modifications, including an additional symmetry-breaking parameter $1/\eta$. 

\begin{eqnarray}
P_{i,i+1}&=&\frac{1}{3}+\frac{1}{\eta}\left(T_{i}^{+}T_{i+1}^{-}+T_{i}^{-}T_{i+1}^{+}\right)+V_{i}^{+}V_{i+1}^{-}+V_{i}^{-}V_{i+1}^{+}\nonumber \\&+&U_{i}^{+}U_{i+1}^{-}+U_{i}^{-}U_{i+1}^{+} +\frac{1}{2}(\lambda_i^3\lambda_{i+1}^3+\lambda_i^8\lambda_{i+1}^8).
\end{eqnarray}
The inclusion of $\eta$ above means we are explicitly breaking the symmetry of the system by changing the energy contribution of turning $\lvert\upa\rangle$ into $\lvert \mda \rangle$ and vice-versa. In Fig.~\ref{fig5} e)-h) we show the result of breaking the SU(3) symmetry (by making $\eta=0.5$) on the dynamics. While the effects in the $\lvert \uparrow \rangle$ and $\lvert \mda \rangle$ components are subtle - an increase in the spin excitation peak as seen in panels f) and g) - in $\lvert \dwa \rangle$ it is more drastic, with the spin contributions being distributed over different low frequencies. The remaining components still preserve isolated peaks for spin oscillations. These results point to the possibility of measuring the independence of spin and density dynamics even in a context where internal symmetries are not perfectly preserved. Naturally, different outcomes for the spin excitations can be expected by choosing a different value of $\eta$, or by breaking the symmetry in a different interaction channel.


\paragraph*{SU(4).}
We now examine the effect of applying our formalism to the case where the number of particles $N$ matches the number of internal components. To that end, we consider the SU(4) fermionic gas with $N=4$ and internal states labeled as $\lvert \upa \rangle$, $\lvert \nea \rangle$, $\lvert \sea \rangle$ and $\lvert \dwa \rangle$. The number of particles in each state is thus given by $N_\upa=N_\nea=N_\sea=N_\dwa=1$ (the so-called 1+1+1+1 infinitely repulsive system with different masses is known to have interesting properties, which were described in \cite{harshmann}). We rewrite the permutation operator for the SU(4) system as

\begin{equation}\label{permusu4}
P_{i,i+1}=\frac{1}{4}+\frac{1}{2}\vec{\lambda}_i\cdot\vec{\lambda}_{i+1},
\end{equation}
where now $\vec{\lambda}$ represents the vector spanning the 15 SU(4) generators \cite{pfeifer2003lie}. In the following, we focus on describing the results only for the $\lvert \upa \rangle$ and $\lvert \nea \rangle$ components.

\begin{figure}
\centering
\includegraphics[width=0.5\textwidth]{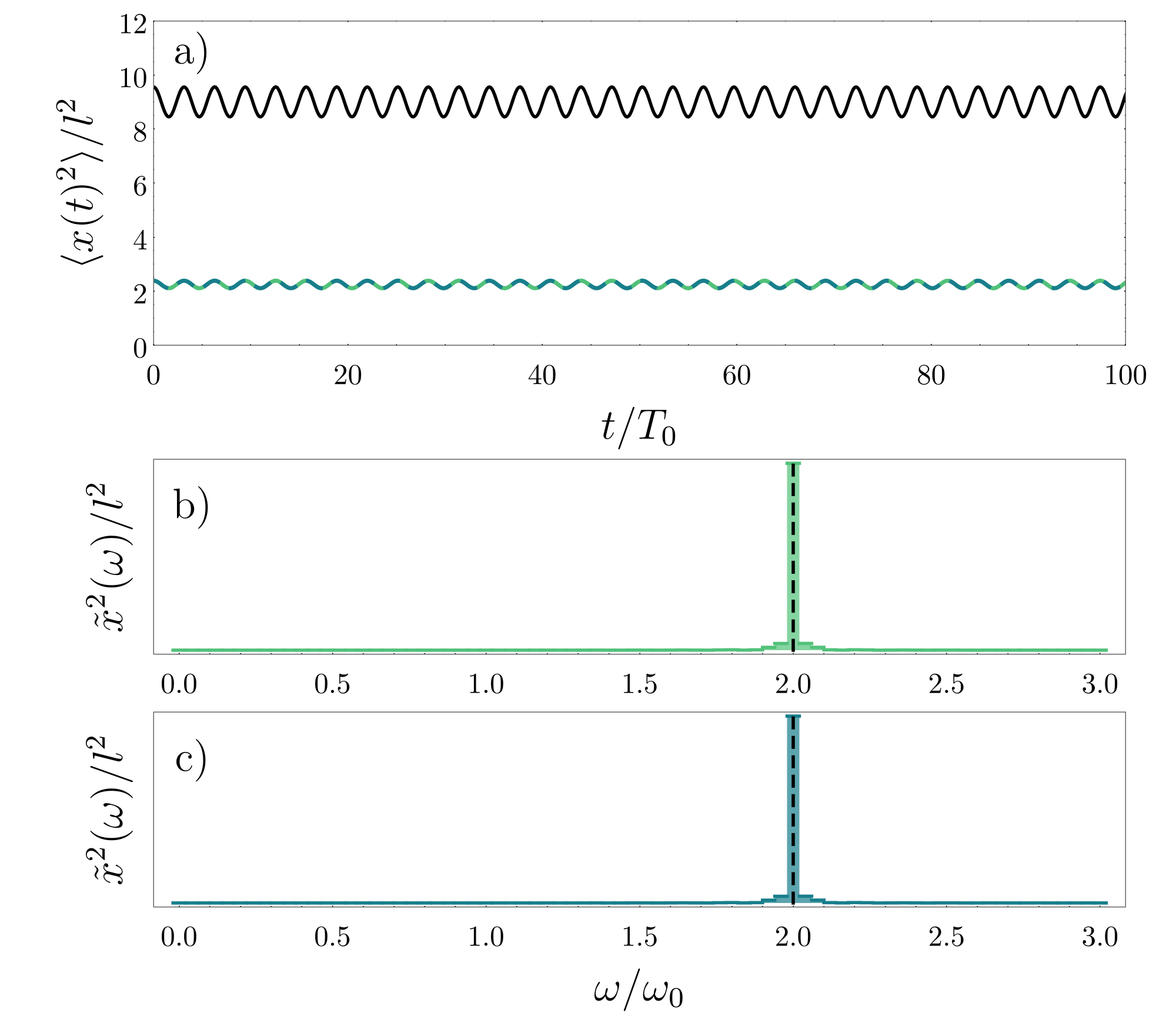}
\caption{Dynamics in the SU(4) fermionic system. a) Time evolution of the squared density widths $\langle x^2_{\upa}(t)\rangle$ (green), $\langle x^2_{\nea}(t)\rangle$ (dashed blue) and $\langle x^2_{\text{c}}(t)\rangle$ (black). Panels b) and c) show the excitation peaks in the frequency domain for $\langle x^2_{\upa}(t)\rangle$ and $\langle x^2_{\nea}(t)\rangle$, respectively. The black dashed lines mark the frequency of the total density oscillations. In these simulations, we assume $g=20$. This figure has been published, with modifications, in Ref.~\cite{reb_thesis}.}
\label{fig6}
\end{figure}

It becomes clear that the behavior $\langle x^2(t)\rangle$ as a function of time is the same for both components shown in Fig.~\ref{fig6} (this also holds for the other two components not shown). This is expected since the number of internal components matches the total number of particles in the system. Moreover, the frequency spectrum shows that the only contributions in the oscillations stem from the density excitations, as opposed to the previous cases. This allows us to interpret the dynamics of the SU($N$) system with strong interactions as the one expected for a gas of impenetrable bosons, as long as the number of particles matches the number of internal components. This conclusion is in agreement with the observation that breathing mode frequencies of the SU($N$) Fermi gas with strong repulsion approaches that of a Tonks-Girardeau gas \cite{fallani}. 

In the models considered here, a vanishing spin signal in the spectrum can additionally be obtained by taking a balanced system with a lower number of internal components (e.g. a 2+2 SU(2) system). This can be explained as a result of the symmetric perturbation to the potential that determines the initial state of the system. Turning off the barrier in this particular case has no effect on the ground state of the spin chain, which remains unchanged even as the density oscillations take place. However, for a matching number of particles and internal components, the results described in this section are the only possible outcomes for a system initialized in the ground state.

\section{Discussion}\label{discussion}
We have presented an analysis of the dynamics of density and spin fluctuations in trapped few-body systems with SU($N$) symmetry. In the limit of strong interactions, the Hamiltonian of this system can be mapped into a spin chain. This mapping allows for a straightforward generalization to systems with more internal components. In addition, it takes into account the geometry of the trapping potential into the set of exchange coefficients of the spin chain. 

The dynamics of the system is obtained after a quench in the trapping potential, where a Gaussian barrier in the center of the harmonic trap is suddenly switched off. This simple protocol is particularly interesting from an experimental point of view, since it requires only minor modifications to the potential, without resorting to spin-selective traps. The sudden change induces the motion of the spatial degrees of freedom, which in turn are reflected in a time-dependence of the exchange coefficients of the spin chain. It is important to point out that, since the system is initialized in the ground state of the spin chain, the motion observed in this sector is only possible due to the quench in the potential. By monitoring the time evolution of the breathing modes given by the oscillations in $\langle x^2(t) \rangle$, we describe the excitation spectrum of SU(2), SU(3) and SU(4) systems. Moreover, in the particular case where the number of internal components exactly matches the number of particles, we see that the spin excitations are completely washed out, and the only contributions are due to density oscillations that agree with those of a spinless Fermi gas. While the theoretical scheme presented here may in principle be hard to generalize to a large number of particles (specially due to the difficulties in calculating the dynamical exchange coefficients over long periods of time), a numerical many-body approach employing the same dynamical protocol may be able to do so. Aside details regarding the true strength of the interactions between non-identical particles, we expect a similar behavior to arise in this context, provided that the same essential ingredients are maintained.

Our results indicate that the decoupling of spin and density dynamics, rather than being exclusively a bulk effect in many-body ensembles, can occur in few-body systems under fairly simple conditions. We point out that multicomponent fermionic systems are the object of current experimental investigation, and the scheme presented in this work could serve as a starting point for experiments with many-body ensembles. The formalism we explored can also be used to predict the behavior of quantum gases with different atomic species (e.g. a bosonic mixture), or generalized to more involved quench protocols, simply by mapping the time evolution of the spatial orbitals into the exchange coefficients of the spin chain under consideration. 

\section{Methods}\label{methods}

\paragraph*{Details for strong interactions.}

The exchange coefficients in Eq.~\eqref{hamper} are determined by the geometry of the trapping potential, and can be calculated as \cite{artem2}
\begin{equation}\label{geo}
\alpha_i=\frac{\int_{x_1<x_2...<x_N-1}dx_1...dx_{N-1}\Big|\frac{\partial \Phi(x_1,...,x_i,...,x_N)}{\partial x_N}\Big|^2_{x_N=x_i}}{\int_{x_1<x_2...<x_N-1}dx_1...dx_N |\Phi(x_1,...,x_i,...,x_N)|^2}.
\end{equation}
where the spinless fermion wave function $\Phi(x_1,...,x_i,...,x_N)$ is built as the Slater determinant of the $N$ lowest occupied orbitals of the potential $V(x)$. The corresponding energy of this state is given by $E_0$ in Eq.~\eqref{hamper}, which is calculated as the sum of the energies of each occupied level. This wave function can be simply viewed as the antisymmetrized version of the Tonks-Girardeau wave function for impenetrable bosons \cite{girardeau}. 
Methods for obtaining the exchange coefficients in different trapping geometries are available and can efficiently calculate $\alpha_i$ for systems with up to $N\sim 30$ \cite{conan,deuretz_mdist}. 

The spin densities given by Eq.~\eqref{spin densities} require calculating the spatial distribution of each individual particle in the trap, which is given by
\begin{equation}\label{onebody}
\rho^i(x)=N!\int dx_1...dx_N \,\delta(x_i-x)|\Phi(x_1,...,x_i,...,x_N)|^2.
\end{equation}
The expressions above involve multidimensional integrals of the spinless fermion wave function $\Phi(x_1,...,x_N)$, which for large systems can become increasingly hard to calculate. Moreover, if we consider time-dependent orbitals $\phi_j(x,t)$, these expressions must be evaluated for all times after the sudden change in the trapping potential. Therefore, we explore the determinant form of the spinless fermion wave function $\Phi(x_1,...,x_N)$ to rewrite these equations in a shape which is suited to calculations in a dynamical context. Equivalent forms of these expressions have been presented in \cite{deuretz1,conan,deuretz_mdist}. The individual one-body densities can be written as
\begin{equation}\label{dens_app}
\rho^i(x)=\frac{\partial}{\partial x}\left( \sum_{j=0}^{N-1} \frac{(-1)^{N-1}(N-j-1)!}{(i-1)!(N-j-i)!j!} \frac{\partial^j}{\partial \lambda^j}\det \left[B(x)-1\lambda \right]\vert_{\lambda=0}\right),
\end{equation}
while the exchange coefficients may be obtained from
\begin{equation}\label{coef_app}
\alpha_i=\sum_{j=1}^N\sum_{k=1}^N(-1)^{j+k}\int_{-\infty}^{+\infty}dx_i \phi'_j(x_i)\phi'_k(x_i)\sum_{l=0}^{N-1-i}\frac{(-1)^{N-1-i}}{l!}\binom{N-l-2}{i-1}\frac{d}{dx_i}\left[\frac{\partial}{\partial \lambda^l}\det \left( B(x_i)-1\lambda \right)_{jk}\vert_{\lambda=0}\right]
\end{equation}
where the matrix $B(x)$ is composed by the single-particle states superpositions $b_{mn}(x)=\int_{-\infty}^{x}dy\,\phi^*_m(y)\phi_n(y)$, and the subscript $jk$ on the right side of Eq.~\eqref{coef_app} indicates that the $j$-th row and $k$-th column are removed. Although we omit the time in these expressions, we consider the single-particle orbitals $\phi_i(x)$ to be time-dependent. 
In Eq.~\eqref{spin densities}, we calculate the spin densities by combining the single-particle spatial distributions with the probability of finding a given spin component at each site. This last quantity can be calculated, for a given spin state $\vert \psi(t) \rangle$, as $m^i_{\uparrow,\downarrow}(t)=\langle \psi(t) \vert (1\pm \sigma^i_z)/2\vert\psi(t)\rangle$.

\paragraph*{Details of the quench protocol.}

The trapping potential at $t=0$ is given by
\begin{equation}\label{doublewell}
V_{t=0}(x)=\frac{1}{2}\omega_0^2 x^2 + V_0 e^{-\left(x/s\right)^2},
\end{equation}
where $V_0$ determines the height of the Gaussian peak and $s$ sets its width. The system is therefore separated in an effective double-well by taking $\omega_0=1$, $V_0=25$ and $s=0.1$. The initial spinless fermion wave function $\Phi(x_1,...,x_N,t=0)$ is constructed with the single particle orbitals obtained by numerical diagonalization, using the $N_s=35$ lowest energy states of the harmonic oscillator. We note that, since the Gaussian peak is large compared to the individual densities of the orbitals, the ground state is quasi-degenerate (the two lowest energy states have nearly the same distribution, with opposite parity).

For $t>0$, we make $V_0=0$ and the spinless fermion wave function is constructed by considering the time-evolved single particles orbitals $\phi_j(x,0)$ according to 
\begin{equation}\label{single_orbitals}
\phi_j(x,t)=\sum_{n=1}^{N_s}c_n e^{-i\epsilon_n t}\psi_n(x)
\end{equation}
where $c_n=\int \psi^{*}(x)\phi(x,0)\,dx$ and $\psi_n(x)$ are the eigenstates of the harmonic oscillator, with $\epsilon_n$ the corresponding eigenvalues. 

\paragraph*{Floquet quasienergy spectrum for a time dependent Hamiltonian}
To obtain the frequency of spin oscillations for the time-dependent spin chain, we focus on finding the Floquet modes for the periodic Hamiltonian and calculating the gaps in the Floquet quasienergy spectrum \cite{floquet}. 
Generally speaking, we are interested in finding the time evolution operator at any time $t$ by solving the following differential equation for $U(t,t_0)$:
\begin{equation}
i\hbar \frac{\partial}{\partial t}U(t,t_0)=H(t)U(t,t_0).
\end{equation}
By obtaining and diagonalizing $U(T,0)$ (where $T$ is the period of the Hamiltonian) we obtain a set of quasienergies $\epsilon_n$ and Floquet modes at $t=0$, which we write as $\lvert u_n(0)\rangle$. The time evolution of the system is then given by
\begin{equation}
\lvert\psi(t)\rangle=\sum_n c_n \exp{(-i \epsilon_n t/\hbar)}\lvert u_n(t)\rangle
\end{equation}
where $\lvert u_n(t)\rangle$ denotes the time-evolved Floquet modes and $c_n=\langle u_n(0)\lvert \psi(0) \rangle$. Once we have calculated $\epsilon_n$, we can easily obtain the energy gaps that contribute to the time evolution of the system by considering the dominant contributions given by $c_n$. In a small system, this is a fairly straightforward process, since the initial state projects only onto a few Floquet modes.

\bibliographystyle{naturemag}
\bibliography{biblio}

\section*{Acknowledgements}
Parts of this work have been published, with modifications, in Ref. \cite{reb_thesis}.
The authors thank Xiaoling Cui, Xi-Wen Guan and David Petrosyan for their useful comments on the manuscript. Leonardo Fallani, Jacopo Catani and Artem Volosniev are also thanked for interesting discussions on spin-charge separation. The following agencies - Conselho Nacional de Desenvolvimento Científico e Tecnológico (CNPq), Coordenação de Aperfeiçoamento de Pessoal de Nível Superior (CAPES), the Danish Council for Independent Research DFF Natural Sciences and the DFF Sapere Aude program - are gratefully acknowledged for financial support. 

\section*{Author Contributions}
The project development, calculations and writing of the manuscript were performed by R.E.B under the supervision of A.F. and N.T.Z.

\section*{Additional Information}
The authors declare no financial and/or non-financial competing interests.
\end{document}